\documentclass[letterpaper]{article} % DO NOT CHANGE THIS
\usepackage{aaai2026}  % DO NOT CHANGE THIS
\usepackage{times}  % DO NOT CHANGE THIS
\usepackage{helvet}  % DO NOT CHANGE THIS
\usepackage{courier}  % DO NOT CHANGE THIS
\usepackage[hyphens]{url}  % DO NOT CHANGE THIS
\usepackage{graphicx} % DO NOT CHANGE THIS
\usepackage{natbib}  % DO NOT CHANGE THIS AND DO NOT ADD ANY OPTIONS TO IT
\usepackage{caption} % DO NOT CHANGE THIS AND DO NOT ADD ANY OPTIONS TO IT
\usepackage{algorithm}
\usepackage{algorithmic}

\usepackage{newfloat}
\usepackage{listings}
\DeclareCaptionStyle{ruled}{labelfont=normalfont,labelsep=colon,strut=off} % DO NOT CHANGE THIS
\floatstyle{ruled}
\newfloat{listing}{tb}{lst}{}
\floatname{listing}{Listing}
\title{AI Unplugged: Embodied Interactions for AI Literacy in Higher Education}
\author {
    Jennifer M. Reddig\equalcontrib\textsuperscript{\rm 1}, 
    Scott Moon\equalcontrib\textsuperscript{\rm 2}, 
    Kaitlyn Crutcher\textsuperscript{\rm 1}, 
    Christopher J. MacLellan\textsuperscript{\rm 1}
}
\affiliations {
    \textsuperscript{\rm 1}Georgia Institute of Technology\\
    \textsuperscript{\rm 2}Independent Researcher\\
    jreddig3@gatech.edu, scottqmoon@gmail.com, kcrutcher3@gatech.edu, cmaclell@gatech.edu
}
\usepackage{bibentry}
\begin{document}

\maketitle

\begin{abstract}
As artificial intelligence (AI) becomes increasingly integrated into daily life, higher education must move beyond code-centric instruction to foster holistic AI literacy. We present a novel pedagogical approach that integrates embodied, unplugged activities into a university-level Introduction to AI course. Inspired by the effectiveness of CS Unplugged in K-12 education, our physical, collaborative activities gave students a first-person perspective on AI decision-making. Through interactive games modeling Search Algorithms, Markov Decision Processes, Q-learning, and Hidden Markov Models, students built an intuition for complex AI concepts and more easily transitioned to mathematical formalizations and code implementations. We present four unplugged AI activities, describe how to bridge from unplugged activities to plugged coding tasks, reflect on implementation challenges, and propose refinements. We suggest that unplugged activities can effectively bridge conceptual reasoning and technical skill-building in university-level AI education.
\end{abstract}

% Uncomment the following to link to your code, datasets, an extended version or similar.
% You must keep this block between (not within) the abstract and the main body of the paper.
\begin{links}
    \link{Materials}{https://github.com/jmreddig/Intro-to-AI-Teaching-Resources}
    % \link{Datasets}{https://aaai.org/example/datasets}
    % \link{Extended version}{https://aaai.org/example/extended-version}
\end{links}

\section{Introduction}

% \textbf{HIGHLIGHT THE GOALS! FIRST-PERSON, UNPLUGGED AI DECISION-MAKING}

% maybe take out the understand-make-reflect cycle? that's what they did but the make/reflect part isn't new, just the understand part

Artificial Intelligence (AI) has become increasingly embedded into everyday life. From social media, to music recommendation, to generative text and images, AI is playing a larger role in how people communicate and make decisions. As AI becomes more ubiquitous, people need to understand how it works, what it can and cannot do, and how to evaluate when a situation calls for an AI solution. This is true especially for those that design, build, and deploy these systems.

However, most student perceptions of AI are shaped not by the technical realities of AI, but by pop culture representations. \citet{hingle2023research} found that student perceptions of AI are formed through everyday interactions and media portrayals, leading to conceptualizing of AI as a magical, superhuman technology. Students rarely enter the classroom with a clear sense of what AI is or how it fits into the broader socio-technical landscape \cite{marx2022brief}. Students may have a hard time reconciling their understanding of AI through media and the routine AI algorithms they learn to code in class. As such, many students are learning to build systems in computer science classes they do not fully understand.

Many introductory AI courses emphasize algorithmic techniques, mathematical formulas, and code-heavy assignments. Instruction is often delivered through lectures, and critical analysis of AI is reserved for the end of the semester, if time allows. K-12 AI education frameworks, such as the framework from the United Nations Educational, Scientific, and Cultural Organization \cite[UNESCO;][]{miao2022k} or the framework proposed by \citet{ng2021conceptualizing}, emphasize the need to go beyond simply building AI systems. University education should pick up where K-12 education leaves off and also teach fundamental AI concepts, discuss whether and when an AI solution is appropriate, and explore the ethical implications of AI use.

% Why is code-level understanding not enough? 

% The United Nations Educational, Scientific, and Cultural Organization (UNESCO) \cite{miao2022k} examined AI curriculum and found three main domains that encompass K-12 AI curriculum - AI foundations, using and developing AI systems, and ethics and social impact. To gain a holistic understanding of AI concepts, students need to not only build AI systems, but know when AI is an appropriate solution, what the potential social and ethical implications are for an AI solution, and have an abstract understanding of how AI functions. Ng et. al, 2021 \cite{ng2021conceptualizing} arrived at a similar framework for teaching AI literacy: know and understand, use and apply, evaluate and create, and ethical issues.  Lao, 2020 \cite{lao2020reorienting} used the categories of knowledge, skills, and attitudes, also highlighting the need for students to go beyond `doing'. University AI education centers around `use and apply' and `using and developing AI systems', but lacks the other components of these educational frameworks. University education should pick up where K-12 education leaves off and further student learning toward a holistic understanding of AI.

% stuff about embodied cognition and requirements of CS unplugged

To address potential gaps in university-level AI education, we developed a curriculum inspired in part by the CS Unplugged concept used in K-12 settings. In particular, our unplugged activities emphasize embodied cognition, the theory that human cognition is shaped through physical, interactive experiences \cite{varela2017embodied}. \citet{Abrahamson_Lindgren_2014} say that ``conceptual reasoning originates in physical interaction and becomes internalized as simulated actions.'' Our curriculum focuses on using embodied cognition to give students a first-person experience of AI decision-making to support their conceptual understanding of these complex algorithms. We provide several opportunities for students to simulate reasoning as AI agents.

We offer a potential model for teaching AI in ways that are both rigorous and reflective, preparing students to build intelligent systems with a deep understanding of their complex algorithmic underpinnings. In this paper, we:
\begin{itemize}
    \item Present four unplugged activities for teaching AI,
    \item Describe how to bridge from unplugged activities to plugged coding tasks,
    \item Probe the viability of using an unplugged curriculum for AI in higher education, and
    \item Identify opportunities and challenges for unplugged instruction at the university level.
\end{itemize}
We describe a sample implementation of a joint unplugged/ plugged Introduction to Artificial Intelligence course taught during the summer 2025 semester. We reflect on student interactions and provide suggestions for further iterations.

% We present several unplugged activities to teach AI concepts by giving students a first-person perspective on AI decision-making and describe how to incorporate them into higher education AI curriculum. 

\section{Background}

AI literacy, as defined by \citet{long2020ai}, is a set of competencies that enable people to communicate, work, and live with AI technologies. These competencies involve understanding different definitions of intelligence, how computers make decisions, and how training data informs the results from AI algorithms. They also provide some design considerations for supporting learner-centered AI education. While all of the design considerations are important for building a holistic understanding of AI, we focused on three design considerations when building this course. \textit{Embodied Interactions} give students a first-person perspective of how AI agents reason and make decisions. The \textit{Critical Thinking} design consideration encourages students to become critical consumers of AI. The \textit{Low Barrier to Entry} design consideration aims to ease learners' insecurities around math and CS abilities by providing supports so learners can build the necessary prerequisite skills.  

% \textbf{MAYBE SKIP THE PRECONCEPTIONS SINCE WERE NOT ACTUALLY MEASURING AI LITERACY}

% Another design competency is Acknowledging Preconceptions. Learners often come into AI classes with an existing understanding of AI shaped by their daily interactions and pop culture \cite{hingle2023research}. Students commonly anthropomorphize agents and believe that AI has intentions like humans. Media portrayals of AI reinforce the belief that AI is a general intelligence, associated with robotics, or all-knowing and infallible. Students are unaware that AI systems are trained on data and that data quality affects agent outcomes \cite{marx2022brief}. However, Hingle \& Johri \cite{hingle2023research} found that reflective journaling prompts helped students become more aware and critical of AI in their environment. Reflecting on where AI appears every day took AI concepts out of the classroom and helped students see how AI impacts their daily lives. Simple reflection prompts can be all that students need to increase ethical and social awareness of the effects of AI.

Recent research in K-12 best practices for AI education continues to emphasize not only the technical, but also the ethical and conceptual dimensions of AI literacy. \citet{lao2020reorienting} developed the Machine Learning Education Framework with three main outcomes of the curriculum: Knowledge (general knowledge of AI methods), Skills (problem scoping, creating artifacts, analysis of results), and Attitudes (identity, community, self-efficacy).  \citet{ng2021conceptualizing} proposed four instructional goals for AI education: learners should know and understand AI, use and apply AI methods, evaluate and create AI systems, and engage with ethical issues. \citet{miao2022k} (in association with UNESCO) examined these and many more K-12 AI curricula and identified three key domains: AI Foundations like algorithmic thinking and data literacy, Using and Developing AI systems, and Ethics and Social Impact of AI applications within and outside the Information and Communication Technology domain. These frameworks stress the importance of not only understanding AI algorithms and building systems, but analyzing the results and reflecting on ethical  and social issues.

Higher education, however, is increasingly focused on the technological outcomes of AI courses. Game-based and project-based learning are the main approaches for building AI understanding and applying AI methods \cite{ng2023review}. Studies like those conducted by \citet{1611941}, \citet{vargas2020project}, and \citet{de2023using} show that projects improve engagement. These gamified projects promote hands-on application and programming fluency, but students may lose sight of broader concepts as they grapple with implementation details. The current state of AI higher education focuses on the Using and Developing AI Systems and General Knowledge components of AI curriculum frameworks, neglecting ethical reasoning, analyzing and evaluating AI systems, and foundational knowledge like data literacy and bias. As AI becomes more embedded into everyday life, higher education must evolve to include holistic AI education like the K-12 frameworks.

CS Unplugged is a powerful pedagogical tool for making abstract computing ideas concrete and engaging in K-12 computer science classes. These technology-free activities use physical, often collaborative tasks that give students a first-person, embodied experience to support CS understanding. CS Unplugged activities lower the barrier to entry, increase motivation, and build conceptual understanding of CS concepts \cite{battal2021computer, huang2021critical}. Recent research focuses on extending CS Unplugged activities to include AI concepts. AI Unplugged, coined by \citet{lindner2019unplugged}, incorporated AI Unplugged activities into a curriculum to build conceptual understanding. \citet{alraddady2014study} used embodied analogy to teach parallel processing algorithms.  \citet{10.1145/3465074.3465078} proposed an activity to help students understand how agents make connections using a semantic network.  \citet{lim2024unplugged} designed a board game to teach middle school students about facial recognition technology. 
% \citet{song2024artificial} used unplugged activities to teach students how conversational AI agents work. 
\citet{ma2023developing} demonstrates that unplugged activities can be more than a stand-alone activity and are most effective when part of a lesson that transitions to plugged activities. We use CS Unplugged inspired activities to build conceptual understanding of AI techniques.

Unplugged activities pull from the theory of embodied cognition \cite{varela2017embodied}, giving students interactive experiences to support a growing conceptual understanding. \citet{hu2024young} used unplugged activities to build computational thinking skills in young children, demonstrating that the embodied experience facilitated learning. 
% \citet{dai2024integrating} leveraged physical interactions to teach 6th-grade students AI concepts. 
\citet{kwon2025computational} connected abstract concepts to physical experiences through CS Unplugged and mixed-reality. 
\citet{dai2024integrating} supplemented the unplugged approach with plugged activities and reflection components, enhancing embodied analogy to support AI literacy. Embodied classroom experiences can enhance attention, memory, and conceptual understanding \cite{sullivan2018learning}. This perspective aligns with work on embodied interaction, which emphasizes how students’ sensorimotor engagement with learning environments can support conceptual development by surfacing intuitive, action-based ways of thinking \cite{abrahamson2020future, yang2024artificial}.  These embodied experiences allow for a richer mental encoding of information which facilitates improved recall and use of that information, resulting in improved outcomes across subjects like reading, writing, math, and physics \cite{fugate2019role}. We use the same principles to enhance the student experience of learning AI techniques.

Despite the effectiveness of CS Unplugged for engaging and motivating students, unplugged activities are rare in higher education. This research aims to fill the gap. A well-structured unplugged to plugged to reflection learning cycle could turn conceptual understanding into practical skill building and ethical awareness, aligning with the UNESCO framework for holistic AI education. To promote this cycle, our course curriculum incorporates unplugged activities and embodied interactions not only to teach AI concepts, but to support deeper engagement and reflection.

% \begin{figure}
%     \centering
%     \includegraphics[width=1\linewidth]{images/conjecture_mapping.png}
%     \caption{The conjecture map \cite{sandoval2014conjecture} for how the understand-make-reflect cycle will impact student outcomes.}
%     \label{fig:conjecture_map}
% \end{figure}

\section{The Course}

% REDO!!! JUST ONE CLASS, POST-COURSE INTERVIEWS

We present an exploratory case study using embodied, unplugged activities teaching artificial intelligence in higher education. The course, Introduction to Artificial Intelligence, is a general overview course of AI techniques and has four main topics: 1) Search, 2) Reasoning with Uncertainty (Markov Decision Processes and Q-Learning), 3) Probabilistic Reasoning through Time (Dynamic Bayes Nets, HMMs, Filtering), and 4) Machine Learning (Neural Networks, Decision Trees, Classification, Regression). The course is taught over the 12-week Summer 2025 semester.

% At the start of the semester, \_\_\_ students were enrolled. By the end of the semester, \_\_\_ students were enrolled, for a withdrawal rate of \_\_\_. This is consistent with prior summer semesters which average a \_\_\_ withdrawal rate.

Each of the main topics has an in-depth coding project. Students use Jupyter Notebooks to 1) develop algorithms to find a sequence of actions to navigate a house and find treasure, 2) approximate an optimal policy through repeated trials, 3) localize a hidden object, and 4) build a neural net from the ground up. After each coding project, students reflect on the applied algorithms' uses, benefits, and potential issues through Socratic Mind \cite{hung2024socratic}. This assignment structure is consistent with the Fall and Spring offerings. 

During the summer semester, we modified the in-class lecture time to focus on unplugged experiences. One activity was conducted each class session, sometimes repeating an activity multiple times over several days to add complexity to the base set-up. Following the unplugged experience, students were guided through a plugged reconstruction of the simulation in Python Jupyter notebooks. The in-class instruction explicitly outlined how the unplugged game can be represented as an AI problem, transitioning from unplugged intuition to mathematical formalization, building toward the practical implementation in each project.

\section{Activities}

Our design goals for these activities were a combination of the defining characteristics of unplugged activities and more specific goals derived from the context and content of the course. Like any unplugged activity, students do not use computers, and instead use physical objects with paper and pencil. There is special emphasis on interaction with other students in a game-like atmosphere to encourage playing with concepts collaboratively. Where possible, we used familiar mechanics like dice and cards to foster a sense of play, to capitalize on students' prior experience with probabilistic reasoning, and to emphasize that probabilities involved in a problem are not necessarily arbitrary parameters, but can be an intrinsic part of the problem or application.

Because these activities are AI specific, we wanted to align students' decisions throughout the activities with the decision making process of AI agents. By highlighting first-person decision making, the activities expose the limitations an AI agent faces. They allow students to experience the steps of an algorithm for themselves and intuit how each step fits into the broader process. We frequently use hidden information to encourage step-by-step decision making instead of guessing a solution from a birds-eye view. 

As tools for a university-level course, we wanted these activities to transition well into a more technical discussion. The activities are designed to fit neatly into the mathematical formalism of the broader problem, and they are simple enough to be solved by hand without being trivial. By using the activity to motivate the formal problem statement, the mathematics should feel more natural as well.

The activities are designed to be fully integrated into the course. After each activity, we had students articulate their reasoning and strategies in order to explicitly connect the ways students make decisions to the ways AI make decisions. Then we followed up with a mathematical formalism, now motivated by the activity and their own experience. Finally, students constructed an AI agent that uses the formalism to make decisions using Python, transitioning from unplugged to plugged.

% Most artificial intelligence methods are based on human intelligence, and students are already experts in their own decision-making. We use physical manipulatives, metaphors, and games to put students in a situation where they need to make an optimal decision under stochasticity or uncertainty, such as card or dice games. Through these activities, the world dynamics become clear and students can reason concretely, even when outcomes are uncertain. By having students articulate their reasoning and strategies, we can make explicit connections from the ways students make decisions to the ways AI make decisions. When we follow each activity up with a mathematical formalization of how AI would represent and process the problem, students can see how and why we formalize phenomena in particular ways, instead of seeing arbitrary definitions before grasping the big picture. Students then construct an AI agent that uses the method to make decisions using Python, transitioning from unplugged to plugged.

We present in this paper four activities, but this is not an exhaustive list. We chose these activities to represent the broad scope of AI topics that could be taught using unplugged instruction. Complete instructional materials and additional activities are available.

% ONE MORE PART HERE ABOUT THE EXPLICIT REFLECTION AND CONNECTION DRAWN, LIKE THE QUOTE BELOW

% abrahamson - "Embodied designs will more effectively lead to conceptual development if students are asked to articulate their strategies for interacting with materials in the environment"

\subsection{Search}

How does an agent know what actions to take to achieve a goal? We designed an activity, \textit{Becoming Search}, that puts students in a first-person perspective and utilizes distributed information to search for a goal state. Search algorithms are typically taught from a birds-eye view, where all connections and state information are available at once. An AI agent using a search algorithm does not have the same perspective.

% This activity is designed to simulate the scenario-agnostic nature of search algorithms and give students a first-person perspective of the information available in each state. Students are used to seeing state-space graphs and search trees from a birds-eye view, where all connections and state information is available at once. 

In prior semesters, we have observed that students often do not understand how an agent can make decisions only seeing one state at a time, when they, as humans, can see all states and connections in the search graph. Students also often misunderstand the value of a heuristic as an estimate and instead try to compute the actual distance of a state to the goal, rather than using local state information to estimate the distance. This activity is designed to combat these misunderstandings before students try to code an agent.

\begin{figure}
    \centering
    \includegraphics[width=\linewidth]{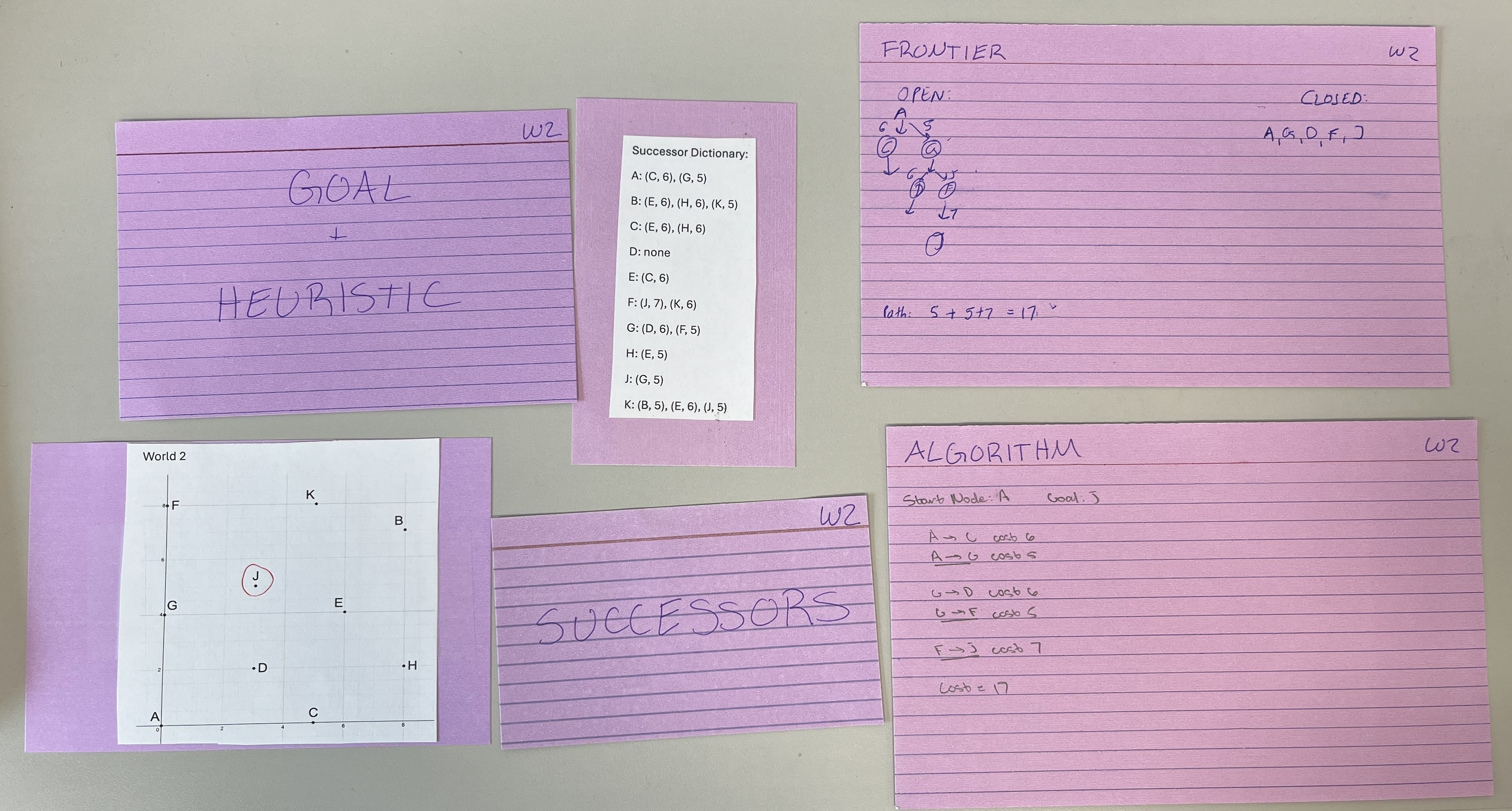}
    \caption{Sample cards that are given to each role with distributed information. Students use the cards to share information and trace through the search scenario.}
    \label{fig:search_example}
\end{figure}

Students form groups of four. Each student in the group has a role. Each role has access to a subset of information about the world, represented on index cards like in Figure \ref{fig:search_example}. The roles are:
\begin{itemize}
    \item \textbf{The Successor Dictionary} - This role has access to all the connections in the state space graph in the form of a dictionary. When asked about a particular state, this role looks up the state in the dictionary and reports all of the connections. 
    % \begin{itemize}
        % \item This role has access to all the connections in the state space graph in the form of a dictionary. When asked about a particular state, this role looks up the state in the dictionary and reports all of the connections. 
        % If the scenario includes weighted edges, the weights are included. If the connections are specific actions, the name of those actions are included (e.g., `pick up the coin', or `open the door').
    % \end{itemize}
    % \begin{itemize}
    %     \item Input: a state name/id
    %     \item Output: all the successors of that state and the actions/costs to get there
    % \end{itemize}
    \item \textbf{The Goal Test} - This role is aware of goal conditions. If the goal is a single state, this role knows the state id. If the goal is a set of state information (e.g., `You win when you are within 10 miles of Budapest.' or `The key is in your inventory and you returned to the start.'), this role processes each state to see if the goal conditions have been met.
    % \begin{itemize}
        % \item This role is aware of goal conditions. If the goal is a single state, this role knows the state id. If the goal is a set of state information (e.g., `You win when you are within 10 miles of Budapest.' or `The key is in your inventory and you returned to the start.'), this role process each state to see if the goal conditions have been met.
    % \end{itemize}
    % \begin{itemize}
    %     \item Input: a state name/id
    %     \item Output: Yes or No, if that state meets the goal conditions
    % \end{itemize}
    \item \textbf{The Frontier} - This role uses a stack of index cards or handwritten notes to keep track of states to be visited. It inserts new states into the stack and when prompted, tells the Algorithm which state to process next.
    \item \textbf{The Algorithm} - This role orchestrates the algorithm's functionality by utilizing all the other roles. They keep track of the visited states and parent states. This role follows the pseudocode of the algorithm and calls upon the other roles to search for the goal. 
    % \begin{itemize}
        % \item This role orchestrates the algorithm's functionality by utilizing all the other roles. They keep track of the visited states and parent states. This role follows the pseudocode of the algorithm and calls upon the other roles to search for the goal. 
        % \item Input: a starting state name/id
    % \end{itemize}
\end{itemize}

Students start with an initial state, and the Algorithm uses the Successor Dictionary to populate the Frontier. The Frontier starts by using a queue, adding new states to the end and pulling the next state from the front. The Goal Test evaluates each state from the Frontier until a goal state is found. Students then swap roles and repeat the exercise on a new scenario so each student gets to experience every role.

The initial setup models Breadth-First Search, but simple changes in one role's function can change the process as a whole. To model Depth-First Search, change the function of the Frontier from using a queue, to using a stack (add new states to the end and pull the next state from the end). To highlight the importance of action sequences, have students report not just the states passed through to get to the goal, but the actions taken from each state. In this variant, the Successor Dictionary has not just the connections, but the names of those connections, such as `go left', `go right', `go down stairs', etc. To model Uniform-Cost Search, modify the information given to the Successor Dictionary to include action costs, and have the Frontier use a priority queue (order the states by accumulated costs). To model Greedy Search or A* Search, modify the Goal Test role to also estimate the distance a state is from achieving the goal by specifying a heuristic and have the Frontier order states using the estimate. With this simple setup, students can see how minute changes in components allow the algorithm to handle different scenarios, without changing the algorithm's process.

% Within the student groups, work through different scenarios, starting from the traditional abstract view of search (i.e., find a path from S to G, state information and actions are abstracted) and slowly adding in complications like action names, goal conditions, and state information. Change the algorithm by changing the Frontier's behavior. Add path weights and the Heuristic role to simulate informed search algorithms. This activity is designed to be repeated multiple times on multiple days to support understanding the variety of search algorithms that depend on the same structure, but have different behaviors and solutions based on small component changes.

% \begin{figure}
%     \centering
%     \includegraphics[width=1\linewidth]{images/search_conjecture_map.png}
%     \caption{The conjecture map for how Becoming Search activity will impact student outcomes}
%     \label{fig:search_conjecture_map}
% \end{figure}

\subsection{Markov Decision Processes}

How can an agent make decisions when the outcomes of those decisions are uncertain? We developed a game called \textit{Red and Black Jack} (inspired by Blackjack or ``21'') to introduce students to Markov Decision Processes (MDPs), which formalize planning in stochastic environments. 

% This game, Red and Black Jack, presents students with opportunities to make decisions when outcomes are uncertain. The canonical example for most Markov Decision Processes is GridWorld, where a robot must choose what direction to move in when there is some chance that an unintended action actually executes. Much like Search problems, GridWorld is a birds-eye perspective of decision-making. Red and Black Jack puts students in a first-person perspective where they need to maximize their game winnings at risk of loosing everything. Card games are a familiar environment for most students, and the structure is similar to BlackJack so students can play with very little explanation. Students judge the value of their current `state', and weigh the odds of drawing another card. This setup can be constrained to be similar in size to GridWorld, with the optional added complexity (while still intuitive) of probabilities changing after each action.

Red and Black Jack is a single player game, but the activity breaks students into pairs so they can take turns acting as player and dealer to collaborate on strategies and check scores. Using a standard 52-card deck, setup a `Hit' deck and a `Stand' deck for each pair.  In the `Hit' deck, put two red cards, two black cards, and one face card (the other cards should be non-face cards to distinguish them). In the `Stand' deck, put one red card and one black card. Each turn, the player can either `Hit' or `Stand'. When executing the `Hit' action, the player draws one card from the `Hit' deck, and as long as a face card was not drawn, play continues. Drawing a face card causes the player to `bust', ending the game and rewarding the player with negative points. When executing the `Stand' action, the player draws one card from the `Stand' deck and the game is over. 

At the end of the game, the player calculates their score based on the maximum number of cards of any single color. If the player drew a face card, they `Bust' and earn $-5$ points, ignoring any other cards. If the player took the `Stand' action and has at most $1$ black or red card in their hand, they earn a `Single' and score $1$ point. If the player took the `Stand' action and has at most $2$ black cards or $2$ red cards, they earn a `Double' and score $5$ points. If the player took the `Stand' action and has at most $3$ red or black cards, they earn a `Triple' and score $15$ points. If the player manages to `Hit' and draw every card \textit{except} the face card, they score a `Jackpot' and earn $30$ points. 

\begin{figure}
    \centering
    \includegraphics[width=\linewidth]{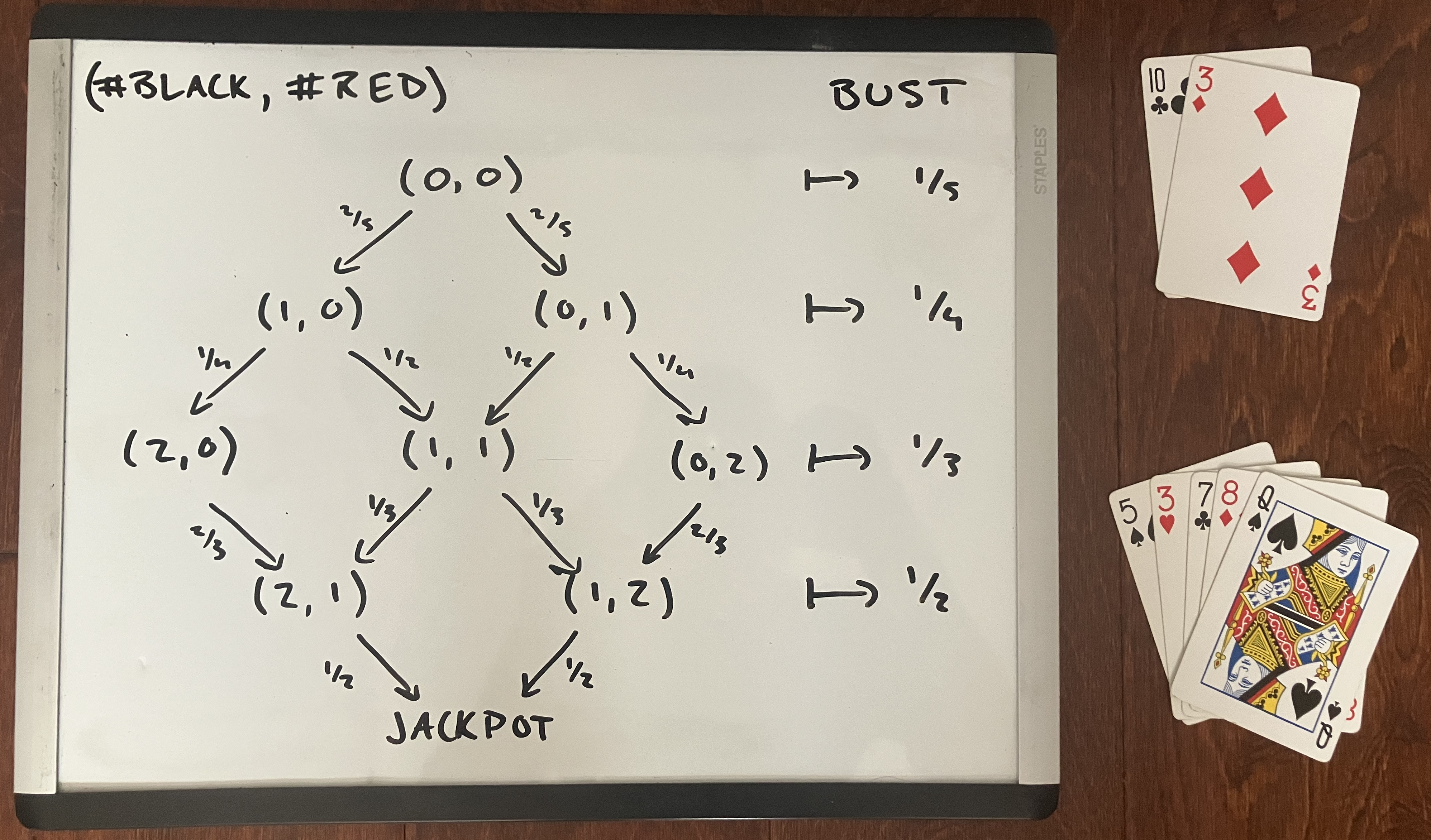}
    \caption{Sample Hit and Stand decks for Red and Black Jack. The ``Hit" transitions for MDP that represents this game can be expressed in this grid format.}
    \label{fig:red_and_black_jack_example}
\end{figure}

Students can play multiple rounds and build an intuition for what they should do in each situation, and then discuss how to formalize the game as an MDP. In our formalization, we had 9 non-terminal states (the player has drawn 0 red cards and 0 black cards, or 1 red card and 2 blacks cards, etc.) and 5 terminal states (`Bust', `Single', `Double', `Triple', and `Jackpot') as visualized in Figure \ref{fig:red_and_black_jack_example}. Playing through the game eases the discussion of filling in the transition probabilities and the notion of a policy as a solution to the MDP. The size of the problem is quite small and has built in symmetry (having 2 red, 1 black is functionally the same as 1 red, 2 black), so students can perform value iteration by hand with relatively few computations. The algorithm converges within a few iterations, and we chose the rewards so the end values would be ``nice" numbers.

Following formalizing MDPs, students can explore how different changes in setup can result in differing policy behaviors. For example, significantly increasing the value of `Jackpot' will encourage more risky behavior, while a higher cost to drawing the `Bust' card or having more `Bust' cards will make the optimal policy appear more cautious. Students can make theoretical predictions about policy changes, then enact them by simulating the game in code or performing value iteration.

% \begin{figure}
%     \centering
%     \includegraphics[width=1\linewidth]{images/mdp_conjecture_map.png}
%     \caption{The conjecture map for how Red and Black Jack will impact student outcomes.}
%     \label{fig:mdp_conjecture_map}
% \end{figure}

\subsection{Q-Learning}

How can an agent make decisions from rewards without prior knowledge of the transition probabilities? We designed an activity, \textit{Q-Maze}, that uses each student as one pass of a q-learning agent through the environment. Collectively, students simulate the repeated trials of an agent and watch as the q-values distribute through q-table. Students often remember how to \textit{use} a q-table (pick the action with the highest estimated future reward), but can struggle trying to \textit{build} a q-table (picking the maximum valued action when all the q-values are zero may not be helpful). In addition, students often confuse the q-value, the agent's future reward estimate, with a reward which is given from the environment. This first-person perspective is designed to help students differentiate between the two as well as understand how actions should be chosen before the q-table is filled up.

\begin{figure}
    \centering
    \includegraphics[width=\linewidth]{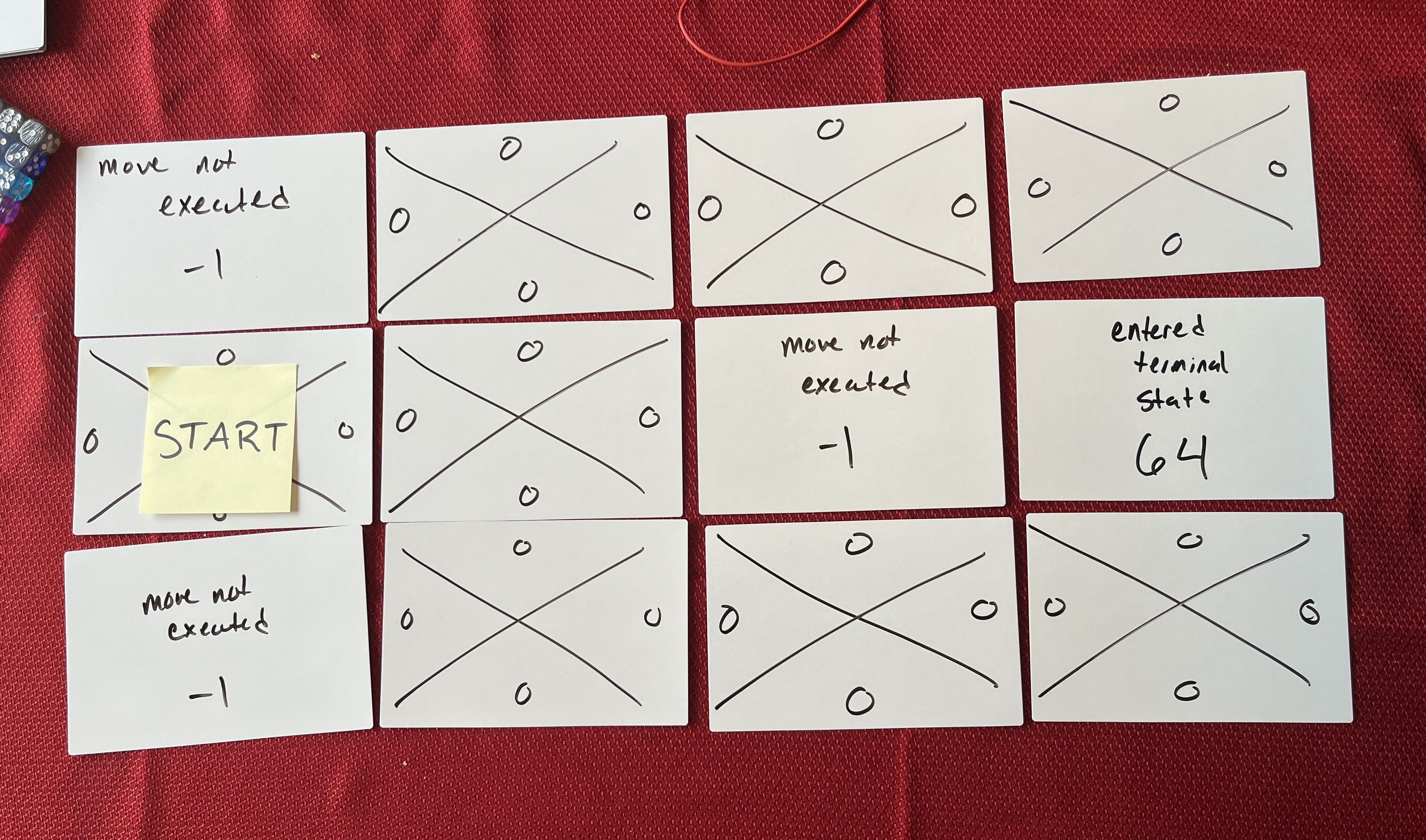}
    \caption{A sample GridWorld environment for Q-Maze.}
    \label{fig:q_learning_whiteboard}
\end{figure}

Start by setting up a grid on the floor (3x3, 4x4, 3x4, 5x5, etc as space allows). Each cell represents one state in GridWorld. Each state has a piece of paper or whiteboard on top of it. In the center of the paper is the `reward' for this particular state. Along the outside of the paper are the actions available in each state and the q-value for each of those actions, as show in Figure \ref{fig:q_learning_whiteboard}. Each student makes one pass through the maze, with the number of allowable actions being the length of the grid plus the width of the grid. One `episode' through the grid looks like:

\begin{itemize}
    \item Flip over the whiteboard for the starting position to see the available movement options and anticipated reward.
    \item Roll a dice to see if you explore and pick a random action, or exploit the existing q-table and take the action with the maximum q-value
    \item Flip over the whiteboard for the chosen action and look at the actual reward
    \item Use the Bellman Equation to update the q-value of the action taken.
    % \begin{itemize}
    %     \item This is the Bellman Update, may adjust to the actual equation, but would introduce more complication to the live updates
    % \end{itemize}
    \item Flip back over the previous state
    \item Repeat until reaching a terminal state or using up all actions
\end{itemize}

Each student should experience navigating through the grid when the q-values are all zeroes, and navigating when the q-values have mostly converged. This requires multiple grids for students to navigate at different points in time. Through trials in early states and in late stages, students will see the need for exploration in early trials and exploitation of the learned q-values in later trials. Once all students have explored the grids, flip over all the whiteboards and lead a discussion on the Q-table that the class converged to, how their individual experiences lead to a more complete picture of the optimal action in each state, and how to translate the converged q-values into an action policy. As an extension, you could modify the algorithm parameters, adjusting how frequently students explore/exploit or how much of the future reward transmits, or what percent of the original estimate is kept on each update. Students can feel the impact of each parameter by repeating the activity, or run their own experiments in a code implementation.

% Ideally, there are multiple grids, and students who explore one grid early on when the q-values are zero will have the opportunity to explore another grid that has been populated. 

% Once all students have explored the grid, flip over all the whiteboards and discuss the Q-table that the class converged on, and how to translate that into an action policy. Reinforce how this models Q-learning, learning through trials and sampled experience. They either explored new states or exploited known pathways. Brainstorm how to build an agent that balances exploration with exploitation.

% \begin{figure}
%     \centering
%     \includegraphics[width=1\linewidth]{images/q_learning_conjecture_map.png}
%     \caption{The conjecture map for how the Q-learning activity will impact student outcomes.}
%     \label{fig:q_learning_conjecture_map}
% \end{figure}

\subsection{Hidden Markov Models}

How can an agent determine its state given noisy sensor data? Up until now, an agent's state has always been known deterministically. We designed an activity, \textit{Two Spies}\footnote{Inspired by the mobile game Two Spies}, to introduce students to Hidden Markov Models (HMMs). Students must hunt down and capture a spy roaming the country in deep cover. The players collect a series of observations after each action, which they put together into an informal `belief' model to track down the most-likely true location of the hidden spy. The deep-cover spy simulates a Hidden Markov Model. Students roll dice to determine what state to transition into and again to determine what the observations report. Students get hands-on experience simulating both the transition model and the sensor model of an HMM. 

During game play, the Deep Cover Spy transitions then reports an observation. Through play-testing, we found that making the observations refer to regions rather than individual cities led to more interesting inferences. The Deep Cover Spy records their actual state and the observation. The Hunter Spy then picks an action - Move, Stay, or Capture. The Hunter Spy moves deterministically to an adjacent state, stays in the current state, or attempts to capture the Deep Cover Spy in the current location. The Hunter Spy records the observation and their chosen action. Given the small state space, the players are given six rounds to either capture the opposing player or evade capture. Students play the game in pairs, swapping roles after three games.

\begin{figure}
    \centering
    \includegraphics[width=\linewidth]{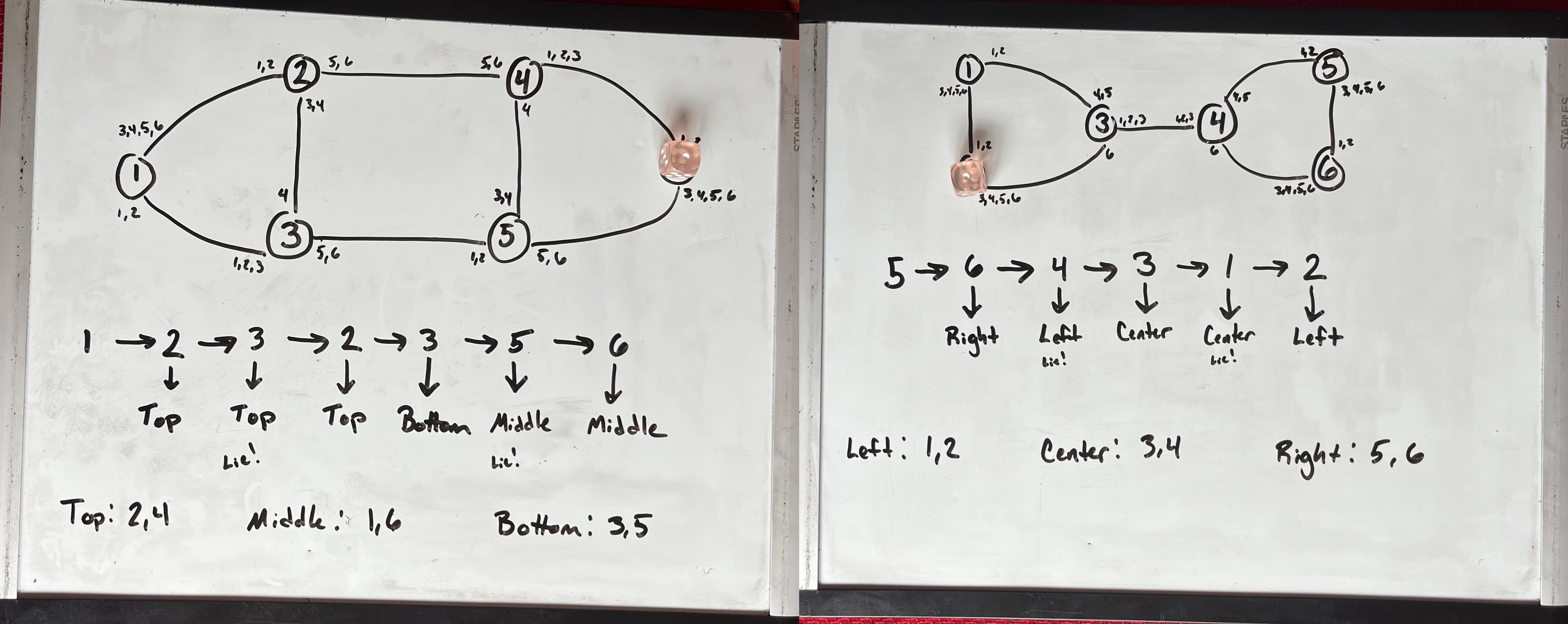}
    \caption{Two sample maps with the HMM transitions and observations, from the perspective of the Deep Cover Spy that simulates the HMM.}
    \label{fig:hmm_whiteboard}
\end{figure}

After playing several rounds in both roles, students can formalize the HMM using mathematical notation. They define the transition matrix and sensor matrix and work through each discrete inference to see how these mathematical calculations reinforce their informal belief. They can also explore how the belief informs sequential decision-making as they build a policy for when to execute each action. After writing the exact inference algorithm in code, students can recreate and reassess their previously played games.

\section{Reflection}

\subsubsection{Successes}

One of the most notable successes was high student engagement. Though we did not require attendance as part of the course grade, class sessions routinely attracted 75-80\% of students through the end of the semester. In prior semesters with more lecture-based instruction, less than 25-30\% of students still attended class sessions at the end of the semester. When students were in class, they were discussing loudly with each other their wins, decisions, and strategies. Every student offered up insight into their discoveries either verbally during the class-wide debrief or through text submission of their personal reflection. 

When converting the unplugged simulation to a technical formalization of the problem, numbers that may appear as arbitrary choices when using an example like GridWorld now appear as obvious properties of the problem. Students could apply their familiarity with dice and card probabilities to decision-making under uncertainty. More than just being familiar with game mechanics, students could translate their own thinking and decision-making processes into algorithms that an AI agent uses. 
% \begin{quote}
%     The strength of Unplugged approaches particularly comes into effect: the students can relate to previous knowledge about how their own thinking works when approaching the topic of artificial intelligence. They observe or compare how certain mechanisms of thinking are automated and how machine learning is realized so that the process of creating artificial intelligence can be retraced. Reflecting on their own thinking and ways of acting makes it easier for the students to fathom out the topic.
% \end{quote}

\citet{lindner2019unplugged} noted that connecting student thought processes to the ways a machine learns makes it easier for K-12 students to comprehend artificial intelligence. This finding appears to hold in higher education. Students were able to reason about each scenario and design rough approximations of the algorithms used, often without prompting. Across all activities students linked their intuitions to formal solutions. For example,
\begin{itemize}
    \item Groups collaborated and shared information to search for a path to a goal, then coded helper functions that mimicked each role in Becoming Search.
    \item Students quickly grasped the explore-exploit trade-off in Q-Maze, hoping for the “explore” dice roll when the Q-table was full of zeros and preferring to pick the maximum-valued action once more entries were filled in. They then used their experience to tune parameters in their Q-learning coding project.
    \item Students expressed joy at seeing their predicted optimal policy emerge by following the value iteration algorithm in Red and Black Jack, and explained how to modify the MDP to encourage different policy behaviors.
    \item Students spontaneously used the extra dice as tokens to represent their belief states, narrowing down the possibilities until only one state was plausible in Two Spies.
\end{itemize}
When we followed each activity with both a manual walkthrough and a coding session, translating the activity into a computer-readable algorithm seemed obvious, since the scenario constraints were familiar properties of the game. Students were able to reason about the algorithm well enough to create a technical implementation in another scenario.

\subsubsection{Challenges}

One of the challenges of a summer semester is that the course must be taught in twelve weeks instead of sixteen. As a survey course, Introduction to Artificial Intelligence exposes students to a variety of methods and losing four weeks reduces the amount of content we were able to cover. In addition, one of the main barriers to introducing active learning techniques is that professors can achieve greater content coverage via traditional lecture than by using active learning techniques \cite{petersen2020tyranny, bonwell1991active}. Running these activities did require extra time for setup, exploration, discussion, coding, and reflection. However, by spending more time on each activity, students were able to connect hands-on experiences to formal algorithmic representations, articulate the reasoning behind computational decisions, and retain key ideas beyond the specific examples presented in class. Judging a course only by how much content is included may not be the best representation of student takeaways.
% Though the steps involved in each activity --introduce the activity, prepare students to work through it independently, let students explore the bounds of the simulation, discuss how the activity can be modeled with AI techniques, trace the algorithm by hand, replicate the simulation in code, and reflect on algorithmic trade-offs -- were time consuming, students self-reported high engagement, motivation, and confidence derived from participating in these activities. 

In addition, AI courses in higher education often focus on technical details and require a strong foundation in mathematics and programming, barriers to students with math anxiety and low self-efficacy \cite{allen2021toward}. Students need extra support to feel confident and build the prerequisite skills required to succeed in university AI curriculum. For unplugged AI learning to succeed in higher education, it needs to support student confidence and scaffold instruction so students can take an active role in their learning.

Scale is still an issue for unplugged learning in higher education. This course had only 40 students and, while larger than a traditional K-12 class, this roster is significantly smaller than the Fall and Spring semesters which can have up to 300 students in a single section. Orchestrating these activities required significant involvement from the instructor and TAs. Making active learning accessible and easy for instructors to implement is a goal of this case study, and we hope that we have outlined activities that are self-contained and able to be self-managed by student groups.

Finally, the precedent for passive lecture in university course has been long established. Students may resist active learning because it violates their expectations of passive instruction \cite{gaffney2010they}, and has a higher perceived workload \cite{shekhar2020negative}. Despite these perceptions, evidence shows that active strategies produce better learning outcomes \cite{deslauriers2019measuring}. Cultivating genuine student engagement may be a challenge, but it is what allows active learning to have its greatest impact.

% what didn't work well:

% - summer semester already condensed, often spend the entire class session to introduce and play the game, needed extra time to make the transition, arguably the most important part

\subsubsection{Opportunities}

% something also here about the opportunities of higher ed for getting more technical, prior experience in algorithms and typically higher math

One of the biggest advantages to applying an unplugged curriculum at the university level was students' prior experience in algorithms and mathematics. Data Structures and Algorithms was a prerequisite to Introduction to AI, so most students were familiar with stacks, queues, multi-dimensional arrays, and algorithmic thinking. Rather than approaching topics like Breadth-First Search and Depth-First Search as brand-new topics, we demonstrated a new application and perspective of these algorithms. The transition from simulating our distributed perspective of search algorithms with index cards to coding a general version with helper methods was made easier because we could rely on the students' prior knowledge and experience in computer science.

We saw similar benefits when learning about MDPs and HMMs. Students had prior experience at least with the format of matrices, multi-dimensional arrays, and probability. Instead of introducing the concept for the first time, we could focus on refreshing and remediating to build confidence with a familiar data structure. Since students already had foundational probability skills, they could easily reason about the chance of drawing a face card from a diminishing deck or traversing a graph from dice rolls. The instructional team could spend less time working through foundational mathematics and programming concepts and accelerate to more complex technical details while providing individual support and remediation where necessary.

% same thing for HMMs and matrices. they were familiar. we could focus on remediation instead of introduction. get you comfortable with the math instead of teaching the math for the first time.

Most unplugged instruction is aimed at younger students in K-12 courses. As such, instruction focuses on conceptual understanding but can gloss over the deep technical details. Even transitioning to plugged activities in Scratch or Teachable Machine can still hide some of the unique complexities of AI algorithms. Younger students may not have the necessary background to understand highly technical implementation details, but university students often do. Unplugged instruction in higher education has the opportunity to build toward a holistic AI education, from conceptual understanding to the technical underpinnings to implementation from scratch because of the advanced learning and cognitive abilities of university students.

% we could also do the technical details, instead of stopping at an intuition appropriate for the level of development. many cs unplugged activities are geared toward K-12 and younger students who are not ready for technical complexities like the large summations in inferenece by enumeration. but college students are prepared for this, because they've taken at least up to algebra II and potentially many more math and computer science classes. even more, we're picking up where K-12 education leaves off.

\subsubsection{Future Versions}

% For future iterations of this class, we recommend spending more time reflecting on students' activity reasoning, demonstrating how multiple AI methods can be used in each scenario, and repeating the activities over multiple days with increasing complexities.

We found that not all students were willing to vocalize their reasoning during the activity and post-activity discussions. To make the benefits of reflection more accessible, we would ask \textit{all} students to explicitly outline their procedure in writing or verbally. We believe this could help students better compare their informal understanding to the algorithm's formal representation. For example, in the Two Spies activity, we did not fully unpack how students’ inferences based on observations relate to the transition model, nor did we work through the matrix multiplications by hand. In the future, we would walk through these calculations step-by-step to show how the belief probabilities align with their informal guesses of the spy's location at each time-step. We would also have students propose an algorithm for how to combine the transition probabilities with the observation probabilities to update their belief. Asking “What did you do, and how is that similar to what the machine is doing to learn?” can help students recognize parallels between their strategies and the underlying algorithms.

In order to create more continuity between modules, it may be helpful to use multiple AI approaches with each scenario. GridWorld is a powerful example because it is flexible and can be used to demonstrate search algorithms, action stochasticity, sensor stochasticity, and a combination of real-world robot navigation challenges. Our activities can also be solved by a variety of AI methods. For example, to use Q-learning with Red and Black Jack, we could hide the contents of the `Hit' and `Stand' deck and scoring rewards and let students approximate the transition probabilities and rewards through playing the game multiple times. Or to apply Particle Filters to Two Spies, students could use small tokens like M\&M's or Skittles to approximate the particles.

We would also adjust the pacing of the activities by spreading them out across multiple days, allowing us to introduce new components and complexities incrementally. In our current schedule, we ran several variants of the Search activity over two days, which limited the attention students could give to the impact of each change, like the frontier structure, action costs, and heuristics. Extending these over more class sessions would provide space for deeper analysis of each modification. Similarly for the other activities, a multi-day structure would allow students to internalize the mechanics and be better prepared for additional scenario complications. For Red and Black Jack, another session could introduce variants like changing the scoring system or adjusting the composition of the Hit or Stand decks. Students could explore the impact of different parameters, like the future reward discount, the reward structure, or the transition probabilities on the optimal policy. Similarly, for Q-Maze, we would follow-up by adjusting the parameters (like the exploration rate, discount factor, learning rate) to explain the impact of each parameter. Students could play another round of Two Spies attempting to narrow down the true location faster using the inference algorithm, or simulate a particle filter on the map instead of exact inference.

\section{Conclusion}

This case study proposes that integrating embodied, unplugged activities into higher education AI courses can foster sustained engagement and stronger connections between human decision-making and algorithmic reasoning. By giving students a first-person perspective on AI techniques before transitioning to formal mathematical models and coding, we lowered barriers to entry, improved retention of core concepts, and encouraged active participation. This approach gave students a stronger intuitive grasp of core concepts and increased student confidence in applying them. Future work should explore scaling these methods to larger classes, incorporating multi-day activity structures, and applying several AI methods to each activity. Ultimately, unplugged AI education at the university level has the potential to bridge the gap between conceptual reasoning and technical skill, preparing students not just to implement AI systems, but to understand and evaluate them in real-world contexts.

\section{Acknowledgments}
This project is supported by National Science Foundation under Grant No. 2247790 and Grant No. 2112532. Any opinions, findings, and conclusions or recommendations expressed in this material are those of the author(s) and do not necessarily reflect the views of the National Science Foundation.

\bibliography{aaai2026}

\end{document}